\documentclass[12pt]{article}
\usepackage{graphicx}

\renewcommand{\bar}[1]{\overline{#1}}
\newcommand{\beq}{\begin{equation}}
\newcommand{\eeq}{\end{equation}}
\newcommand{\ba}{\begin{eqnarray}}
\newcommand{\ea}{\end{eqnarray}}
\newcommand{\nn}{\nonumber}
\newcommand{\bm}[1]{\mbox{\boldmath $#1$}}
\newcommand{\text}{\hbox}
\newcommand{\slsh}[1]{\mbox{$\not\! #1$}}
\newcommand{\simorder}{\raisebox{-4pt}{$\, \stackrel{\textstyle >}{\sim}\,$}}
\newcommand{\simordertwo}{\raisebox{-4pt}{$\, \stackrel{\textstyle <}{\sim}\,$}}
\begin{document}

\begin{flushright}
SLAC-PUB-9561\\
hep-ph/0211110\\
November 2002
\end{flushright}

\bigskip\bigskip
\begin{center}
{\Large \bf Initial-State Interactions in the Unpolarized Drell-Yan Process
}\footnote{\baselineskip=13pt Work partially supported
by the Department of Energy, contract DE--AC03--76SF00515, and by the
LG Yonam Foundation.}
\end{center}

\vspace{13pt}

\centerline{ \bf Dani\"el Boer$^a$, Stanley J. Brodsky$^b$,
and Dae Sung Hwang$^{c}$}

\vspace{8pt} {\centerline{$^a$ Department of Physics and Astronomy,
Vrije Universiteit, De Boelelaan 1081}}

{\centerline{NL-1081 HV Amsterdam, The Netherlands}}

\centerline{e-mail: dboer@nat.vu.nl}

\vspace{8pt} {\centerline{$^b$Stanford Linear Accelerator
Center}}

{\centerline{Stanford University, Stanford, California 94309,
USA}}

\centerline{e-mail: sjbth@slac.stanford.edu}

\vspace{8pt} {\centerline{$^{c}$ Department of Physics, Sejong
University, Seoul 143--747, Korea}}

\centerline{e-mail: dshwang@sejong.ac.kr}

\vfill \centerline{PACS numbers: 12.38.-t, 12.38.Bx, 13.88.+e, 13.85.Qk}
\vfill

\newpage

\setlength{\baselineskip}{13pt}

\bigskip

\begin{abstract}
We show that initial-state interactions contribute to the $\cos 2
\phi$ distribution in unpolarized Drell-Yan lepton pair production
$p \, p$ and $ p\, \bar p \to \ell^+ \ell^- X$, without
suppression. The asymmetry is expressed as a product of chiral-odd
distributions $h_1^\perp(x_1,\bm{p}_\perp^2)\times \overline
h_1^\perp(x_2,\bm{k}_\perp^2) $, where the quark-transversity
function $h_1^\perp(x,\bm{p}_\perp^2)$ is the transverse momentum
dependent, light-cone momentum distribution of transversely
polarized quarks in an {\it unpolarized} proton. We compute this
(naive) $T$-odd and chiral-odd distribution function and the
resulting $\cos 2 \phi$ asymmetry explicitly in a quark-scalar
diquark model for the proton with initial-state gluon interaction.
In this model the function $h_1^\perp(x,\bm{p}_\perp^2)$ equals
the $T$-odd (chiral-even) Sivers effect function
$f^\perp_{1T}(x,\bm{p}_\perp^2)$. This suggests that the
single-spin asymmetries in the SIDIS and the Drell-Yan process are
closely related to the $\cos 2 \phi$ asymmetry of the unpolarized
Drell-Yan process, since all can arise from the same underlying
mechanism. This provides new insight regarding the role of quark
and gluon orbital angular momentum as well as that of initial- and
final-state gluon exchange interactions in hard QCD processes.

\end{abstract}

\section{Introduction}

Single-spin asymmetries in hadronic reactions have been among the
most challenging phenomena to understand from basic principles in
QCD. Several such asymmetries have been observed in experiment,
and a number of theoretical mechanisms have been suggested
\cite{s90,Liang-Boros,QS-91b,Collins-93b,Anselmino,Boer}.
Recently, a new way of producing single-spin asymmetries in
semi-inclusive deep inelastic scattering (SIDIS) and the Drell-Yan
process has been put forward \cite{BHS,BHS2}.  It was shown that
the exchange of a gluon, viewed as initial- or final-state
interactions, could produce the necessary phase leading to a
single transverse spin asymmetry.  The main new feature is that
despite the presence of an additional gluon, this asymmetry occurs
without suppression by a large energy scale appearing in the
process under consideration.  It has been recognized since then
\cite{Collins-02}, that this mechanism can be viewed as the
so-called Sivers effect \cite{s90,Boer2}, which was thought to be
forbidden by time-reversal invariance \cite{Collins-93b}. Apart
from generating Sivers effect asymmetries, the mechanism offers
new insight regarding the role of orbital angular momentum of
quarks in a hadron and their spin-orbit couplings; in fact, the
same $\vec S \cdot \vec L$ matrix elements enter the anomalous
magnetic moment of the proton \cite{BHS}. The new mechanism for
single target-spin asymmetries in SIDIS necessarily requires
non-collinear quarks and gluons, and in the Sivers asymmetry the
quarks carry no polarization on average. As such it is very
different from mechanisms involving transversity (often denoted by
$h_1$ or $\delta q$), which correlates the spin of the
transversely polarized hadron with the transverse polarization of
its quarks.

In further contrast, the exchange of a gluon can also lead to
transversity of quarks inside an {\em unpolarized\/} hadron. This
chiral-odd partner of the Sivers effect has been discussed in
Refs.\ \cite{Boer,Boer3}, and in this paper we will show
explicitly how initial-state interactions generate this effect.
Goldstein and Gamberg reported recently that
$h_1^\perp(x,\bm{p}_\perp^2)$ is proportional to
$f^\perp_{1T}(x,\bm{p}_\perp^2)$ in the quark-scalar diquark model
\cite{Goldstein:2002vv}. We confirm this and find that these two
distribution functions are in fact equal in this model.  Although
this property is not expected to be satisfied in general,
nevertheless, one may expect these functions to be comparable in
magnitude, since both functions can be generated by the same
mechanism. We investigate the consequences of the present model
result for the unpolarized Drell-Yan process. We obtain an
expression for the $\cos{2\phi}$ asymmetry in the lepton pair
angular distribution.  Here $\phi$ is the angle between the lepton
plane and the plane of the incident hadrons in the lepton pair
center of mass.  This asymmetry was measured a long time ago
\cite{NA10,Conway} and was found to be large.  Several theoretical
explanations (some of which will be briefly discussed below) have
been put forward, but we will show that a natural explanation can
come from initial-state interactions which are unsuppressed by the
invariant mass of the lepton pair.

\section{The unpolarized Drell-Yan process}

The unpolarized Drell-Yan process cross section has been measured
in pion-nucleon scattering: $\pi^- \, N \rightarrow \mu^+ \, \mu^-
\, X$, with $N$ deuterium or tungsten and a $\pi^-$-beam with
energy of 140, 194, 286 GeV \cite{NA10} and 252 GeV \cite{Conway}.
Conventionally the differential cross section is written as
\beq
\frac{1}{\sigma}\frac{d\sigma}{d\Omega} = \frac{3}{4\pi}\,
\frac{1}{\lambda+3} \, \left( 1+ \lambda \cos^2\theta + \mu
\sin^2\theta \cos\phi + \frac{\nu}{2} \sin^2 \theta \cos 2\phi
\right).
\label{unpolxs}
\eeq
These angular dependencies\footnote{We neglect $\sin \phi$ and
$\sin 2\phi$ dependencies, since these are of higher order in $\alpha_s$
\cite{Hagiwara,Gehrmann} and are expected to be
small.} can all
be generated by perturbative QCD corrections, where for instance
initial quarks radiate off high energy gluons into
the final state.  Such a perturbative QCD calculation at
next-to-leading order leads to $\lambda \approx 1, \mu \approx 0,
\nu \approx 0$ at very small transverse momentum of the lepton pair.
More generally, the Lam-Tung relation $1-\lambda -2\nu=0$ \cite{LamTung}
is expected to hold at order $\alpha_s$ and the relation is hardly modified by
next-to-leading order ($\alpha_s^2$) perturbative QCD corrections
\cite{bran93}.  However, this relation is not satisfied by the experimental
data \cite{NA10,Conway}.
The Drell-Yan data shows remarkably large values of $\nu$,
reaching values of about 30\% at transverse momenta of the lepton
pair between 2 and 3 GeV (for $Q^2=m_{\gamma^*}^2= (4 - 12 \, \text{GeV})^2$
and extracted in the Collins-Soper frame \cite{CS}
to be discussed below).  These large values of $\nu$ are not compatible
with $\lambda \approx 1$ as also seen in the data.

A number of explanations have been put forward, such as a higher
twist effect \cite{bran94,Eskola}, following the ideas of Berger
and Brodsky \cite{Berger-80}.  In Ref.\ \cite{bran94} the higher
twist effect is modeled using an asymptotic pion distribution
amplitude, and it appears to fall short in explaining the large
values of $\nu$.

In Ref.\ \cite{bran93} factorization-breaking correlations between
the incoming quarks are assumed and modeled in order to account
for the large $\cos 2\phi$ dependence.  Here the correlations are
both in the transverse momentum and the spin of the quarks.  In
Ref.\ \cite{Boer} this idea was applied in a factorized approach
\cite{Ralst-S-79} involving the chiral-odd partner of the Sivers
effect, which is the transverse momentum dependent distribution
function called $h_1^\perp$.  From this point of view, the large
$\cos 2\phi$ azimuthal dependence can arise at leading order,
i.e.\ it is unsuppressed, from a product of two such distribution
functions. It offers a natural explanation for the large $\cos
2\phi$ azimuthal dependence, but at the same time also for the
small $\cos \phi$ dependence, since chiral-odd functions can only
occur in pairs.  The function $h_1^\perp$ is a quark helicity-flip
matrix element and must therefore occur accompanied by another
helicity flip.  In the unpolarized Drell-Yan process this can only
be a product of two $h_1^\perp$ functions. Since this implies a
change by two units of angular momentum, it does not contribute to
a $\cos \phi$ asymmetry. In the present paper we will discuss this
scenario in terms of initial-state interactions, which can
generate a nonzero function $h_1^\perp$.

We would also like to point out the experimental observation that
the $\cos 2 \phi$ dependence as observed by the NA10 collaboration
does not seem to show a strong dependence on $A$, i.e.\ there was
no significant difference between the deuterium and tungsten
targets.  Hence, it is unlikely that the asymmetry originates from
nuclear effects, and we shall assume it to be associated purely
with hadronic effects.  We refer to Ref.\ \cite{Fries} for
investigations of nuclear enhancements.

We compute the function $h_1^\perp(x,\bm{p}_\perp^2)$ and the
resulting $\cos 2 \phi$ asymmetry explicitly in a quark-scalar
diquark model for the proton with an initial-state gluon
interaction.  In this model $h_1^\perp(x,\bm{p}_\perp^2)$ equals
the $T$-odd (chiral-even) Sivers effect function
$f^\perp_{1T}(x,\bm{p}_\perp^2)$. Hence, assuming the $\cos 2\phi$
asymmetry of the unpolarized Drell-Yan process does arise from
nonzero, large $h_1^\perp$, this asymmetry is expected to be
closely related to the single-spin asymmetries in the SIDIS and
the Drell-Yan process, since each of these effects can arise from
the same underlying mechanism.

The Tevatron and RHIC should both be able to investigate azimuthal
asymmetries such as the $\cos 2\phi$ dependence. Since polarized
proton beams are available, RHIC will be able to measure
single-spin asymmetries as well. Unfortunately, one might expect
that the $\cos 2 \phi$ dependence in $p \, p \rightarrow \ell
\,\bar \ell \, X$ (measurable at RHIC) is smaller than for the
process $\pi^- \, N \rightarrow \mu^+ \, \mu^- \, X$, since in the
former process there are no valence antiquarks present.  In this
sense, the cleanest extraction of $h_1^\perp$ would be from $p \,
\bar p \rightarrow \ell \, \bar \ell \, X$.

\section{Cross section calculation}

In this section we will assume nonzero $h_1^\perp$ and discuss the calculation
of the leading order unpolarized Drell-Yan cross section (given in Ref.\
\cite{Boer} with slightly different notation)
\ba
\lefteqn{
\frac{d\sigma(h_1h_2\to \ell \bar\ell X)}{d\Omega dx_1 dx_2 d^2{\bm
q_\perp^{}}}= \frac{\alpha^2}{3Q^2}\;\sum_{a,\bar a} e_a^2\;\Bigg\{
A(y)\;{\cal F}\left[f_1\overline f_1\right]}\nn\\
&& \mbox{} \qquad
+ B(y)\cos(2\phi)\; {\cal F}\left[\left(2\,\bm{\hat h}\!\cdot \!
\bm p_\perp^{}\,\,\bm{\hat h}\!\cdot \! \bm k_\perp^{}\,
                    -\,\bm p_\perp^{}\!\cdot \! \bm k_\perp^{}\,\right)
                    \frac{h_1^{\perp}\overline h_1^{\perp}}{M_1M_2}\right]
\Bigg\}.
\label{LO-OOO}
\ea
This is expressed in the so-called Collins-Soper frame \cite{CS},
for which one chooses the following set of normalized vectors
(for details see e.g.\ \cite{Boer4}):
\ba
\hat t &\equiv & q/Q,\\
\hat z &\equiv &\frac{x_1}{Q}
\tilde{P_1}- \frac{x_2}{Q} \tilde{P_2},\\
\hat h &\equiv & q_\perp/Q_\perp = (q-x_1\, P_1 -x_2\, P_2)/Q_\perp,
\ea
where $\tilde{P_i} \equiv P_i-q/(2 x_i)$, $P_i$ are the momenta of the two
incoming hadrons and $q$ is the four momentum of the virtual photon or,
equivalently, of the lepton pair.
This can be related to standard
Sudakov decompositions of these momenta
\begin{eqnarray}
P_1^\mu &\equiv & \frac{Q}{2 x_1}\,\bar n^\mu
+ \frac{x_1 M_1^2}{2 Q}\,n^\mu,\\
P_2^\mu &\equiv & \frac{x_2 M_2^2}{2 Q}\,\bar n^\mu
+ \frac{Q}{2 x_2}\,n^\mu,\\
q^\mu &\equiv &\frac{Q}{2}\,\bar n^\mu
+ \frac{Q}{2}\,n^\mu
+ q_\perp^\mu,
\end{eqnarray}
with $Q_\perp^2 \equiv - q_\perp^2 \equiv \bm{q}_\perp^2 \ll Q^2$,
via the identification
of the light-like vectors
\begin{eqnarray}
\bar n^\mu & = & \left[ \hat t^\mu + \hat z^\mu
-\,\frac{Q_\perp^{}}{Q} \hat h^\mu \right], \label{nplusc}
\\
n^\mu & = & \left[ \hat t^\mu - \hat z^\mu
-\,\frac{Q_\perp^{}}{Q}\,\hat h^\mu \right].  \label{nplusc2}
\end{eqnarray}

\begin{figure}[htb]
\centering
\includegraphics[height=2in]{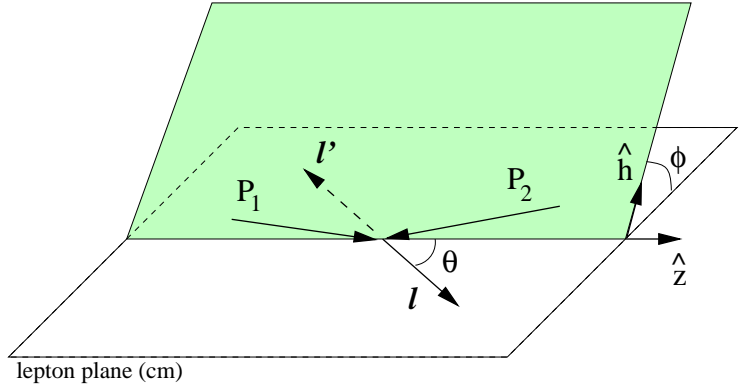}
\caption[*]{Kinematics of the Drell-Yan process in the lepton
center of mass frame.}
\label{DYkin}
\end{figure}

The azimuthal angles lie inside the plane orthogonal to $t$ and
$z$.  In particular, $d\Omega$ = $2dy\,d\phi^l$, where $\phi^l$
gives the orientation of $\hat l_\perp^\mu \equiv \left( g^{\mu
\nu}-\hat t^{ \mu} \hat t^{\nu } + \hat z^{ \mu} \hat z^{\nu }
\right) l_\nu$, the perpendicular part of the lepton momentum $l$;
$\phi$ is the angle between $\bm{\hat h}$ (the direction of
$\bm{q}_\perp$) and $\hat l_\perp$.  In the cross sections we also
encounter the following functions of $y=l^-/q^-$, which in the
lepton center of mass frame equals $y=(1 + \cos \theta)/2$, where
$\theta$ is the angle of the momentum of the outgoing lepton $l$
with respect to $\hat z$ (cf.\ Fig.\ \ref{DYkin}): \ba A(y) &=&
\left(\frac{1}{2} -y+y^2\right) \stackrel{cm}{=} \frac{1}{4}
\left( 1 + \cos^2\theta \right)
, \\
B(y) &=& y\,(1-y) \stackrel{cm}{=}\frac{1}{4} \sin^2 \theta .
\ea
Furthermore, we use the convolution notation
\begin{equation}
{\cal F}\left[f\overline f\, \right]\equiv \;
\int d^2\bm p_\perp^{}\; d^2\bm k_\perp^{}\;
\delta^2 (\bm p_\perp^{}+\bm k_\perp^{}-\bm
q_\perp^{})  f^a(\Delta,\bm{p}_\perp^2)
\overline f{}^a(\bar \Delta,\bm{k}_\perp^2),
\end{equation}
where $\Delta, \bar \Delta$ are lightcone momentum fractions and
$a$ is the flavor index.

In order to obtain the cross section expression one contracts the lepton
tensor with the hadron tensor \cite{Boer,Ralst-S-79}
\beq
{\cal W}^{\mu\nu}=\frac{1}{3} \int dp^- dk^+ d^2\bm{p}_\perp^{} d^2
\bm{k}_\perp^{}\, \delta^2(\bm{p}_\perp^{}+
\bm{k}_\perp^{}-\bm{q}_\perp^{})\, \left.
\text{Tr}\left( \Phi (p) \,
\gamma^\mu \, \overline \Phi
(k) \, \gamma^\nu \right) \right|_{p^+, \, k^-}
+ \left(\begin{array}{c}
q\leftrightarrow -q \\ \mu \leftrightarrow \nu
\end{array} \right),
\eeq
where $p^+=\Delta P_1^+ = \Delta q^+/x_1, \, k^-=\bar \Delta P_2^- = \bar
\Delta q^-/x_2$.  The correlation function $\Phi$ is parameterized in terms of
the transverse momentum dependent quark distribution functions \cite{Boer3}
\begin{eqnarray}
&&\Phi (\Delta , {\bm r}_{\perp};P,S)
\label{ad1}\\
&=&
{M\over 2P^+}{\Big[}f_1(\Delta , {\bm r}_{\perp}){{\slsh{P}}\over M}+
f_{1T}^{\perp}(\Delta , {\bm r}_{\perp})
{\epsilon}_{\mu\nu\rho\sigma}{\gamma}^{\mu}
{P^{\nu}r_{\perp}^{\rho}S_T^{\sigma}\over M^2}
-g_{1s}(\Delta , {\bm r}_{\perp}){{\slsh{P}}{\gamma}_5\over M}
\nonumber\\
&&-h_{1T}(\Delta , {\bm r}_{\perp})
{i{\sigma}_{\mu\nu}{\gamma}_5S_T^{\mu}P^{\nu}\over M}
-h_{1s}^{\perp}(\Delta , {\bm r}_{\perp})
{i{\sigma}_{\mu\nu}{\gamma}_5r_{\perp}^{\mu}P^{\nu}\over M^2}
+h_1^{\perp}(\Delta , {\bm r}_{\perp})
{{\sigma}_{\mu\nu}r_{\perp}^{\mu}P^{\nu}\over M^2}
{\Big]}\ ,
\nonumber
\end{eqnarray}
and similarly for $\overline \Phi$.

We end this section by giving the resulting expression for $\nu$ \cite{Boer}
\beq
\nu = 2\sum_{a,\bar a} e_a^2 \,
{\cal F}\left[\left(2\,\bm{\hat h}\!\cdot \!
\bm p_\perp^{}\,\,\bm{\hat h}\!\cdot \! \bm k_\perp^{}\,
                    -\,\bm p_\perp^{}\!\cdot \! \bm k_\perp^{}\,\right)
                   \frac{h_1^{\perp}\overline h_1^{\perp}}{M_1M_2}\right]\Bigg/
\sum_{a,\bar a} e_a^2 \, {\cal F}\left[f_1\overline f_1\right].
\label{kappa1}
\eeq

\section{Asymmetry calculation}

\begin{figure}[htbp]
\centering\includegraphics[height=2.4in]{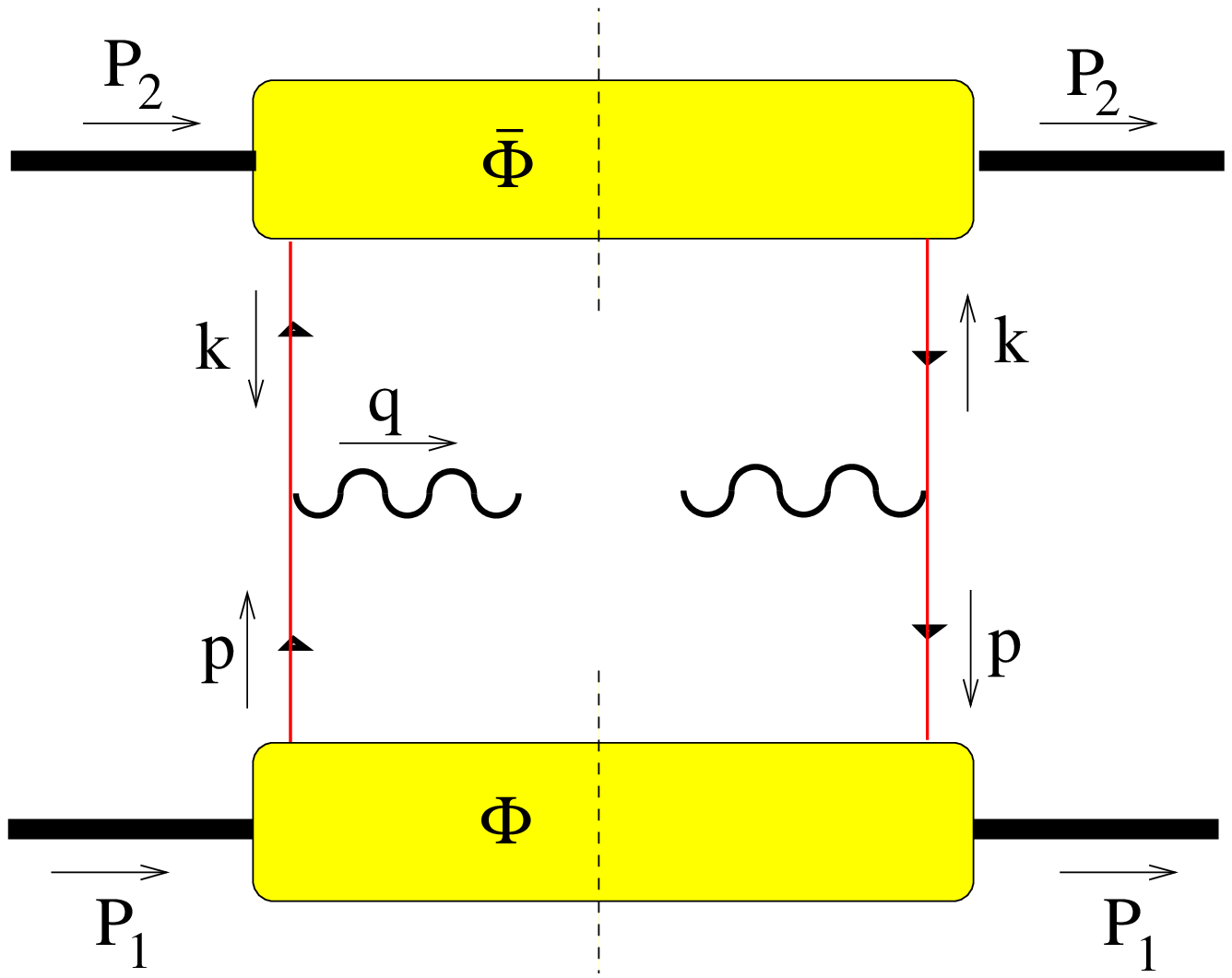} \caption[*]{The
leading order contribution to the Drell-Yan process}. \label{LODY}
\vskip .32in
\centering
\includegraphics[height=2.4in]{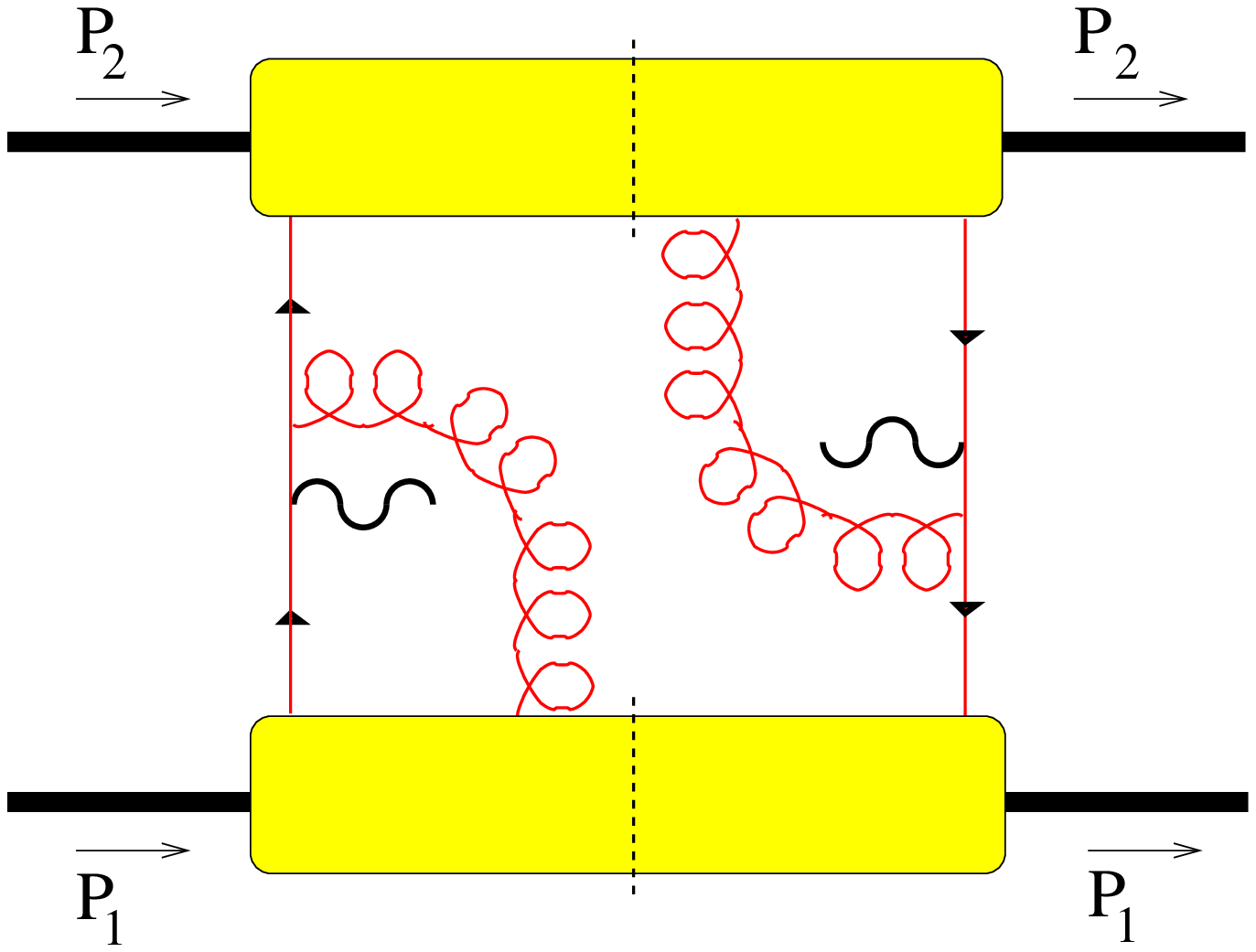} \caption{ The initial-state
interaction contribution to the Drell-Yan process.} \label{LODY2}
\end{figure}

The above cross section in terms of $\Phi$ and $\overline \Phi$ can be
represented by the diagram in Fig.\ \ref{LODY}.

Insertion of the parameterization of $\Phi$ and $\overline \Phi$
will yield the $\cos 2\phi$ asymmetry, among many other terms.
However, in the lowest order quark-scalar diquark model the
diagram Fig.\ \ref{LODY} will not lead to nonzero $h_1^\perp$ in
$\Phi$, and consequently, also not to a nonzero $\cos 2\phi$
asymmetry. To generate such an asymmetry we will include
initial-state interactions corresponding to diagrams such as those
depicted in Fig.\ \ref{LODY2}.  Following the reasoning of Refs.\
\cite{Collins-02,Belitsky}, this should be equivalent to Fig.\
\ref{LODY} with an effective $\Phi$ (and $\overline \Phi$) with
nonzero $h_1^\perp$ function.  Here we do not intend to give a
full demonstration of this in the Drell-Yan process; a generalized
factorization theorem which includes transverse momentum dependent
functions and initial/final-state interactions remains to be
proven \cite{Bodwin}. Instead we present how to arrive at an
effective $\Phi$ from initial/final-state interactions and use
this effective $\Phi$ in Fig.\ \ref{LODY}.  Also, for simplicity
we will perform the explicit calculation in QED. Our analysis can
be generalized to the corresponding calculation in QCD.  The
final-state interaction from gluon exchange has the strength
${|e_1 e_2|\over 4 \pi} \to C_F \alpha_s(\mu^2)$, where $e_i$ are
the photon couplings to the quark and diquark.

The diagram in Fig.\ \ref{LODY2} coincides with Fig.\ 6(a) of
Ref.\ \cite{Jaffe-Ji-91} used for the evaluation of a twist-4
contribution ($\sim 1/Q^2$) to the unpolarized Drell-Yan cross
section.  The differences compared to Ref.\ \cite{Jaffe-Ji-91} are
that in the present case there is nonzero transverse momentum of
the partons, and the assumption that the matrix elements are
nonvanishing in case the gluon has vanishing light-cone momentum
fraction (but nonzero transverse momentum).  This results in an
unsuppressed asymmetry which is a function of the transverse
momentum $Q_\perp$ of the lepton pair with respect to the initial
hadrons.  If this transverse momentum is integrated over, then the
unsuppresed asymmetry will average to zero and the diagrams will
only contribute at order $1/Q^2$ as in Ref.\ \cite{Jaffe-Ji-91}.

First we will calculate the $\Phi$ matrix to lowest order (called
$\Phi_{L}^{\alpha\beta}$) in the quark-scalar diquark model which
was used in Ref.\ \cite{BHS}.  (Although the model is based on a
point-like coupling of a scalar diquark to elementary fermions, it
can be softened to simulate a hadronic bound state by
differentiating the wavefunction formally with respect to a
parameter such as the proton mass.) As indicated earlier, no
nonzero $f_{1T}^\perp$ and $h_1^\perp$ will arise from
$\Phi_{L}^{\alpha\beta}$. Next we will include an additional gluon
exchange to model the initial/final-state interactions (relevant
for timelike/spacelike processes) to calculate
$\Phi_{I/F}^{\alpha\beta}$ and do obtain nonzero values for
$f_{1T}^\perp$ and $h_1^\perp$. Our results agree with those
recently obtained in the same model by Goldstein and Gamberg
\cite{Goldstein:2002vv}.  We can then obtain an expression for the
$\cos 2\phi$ asymmetry from Eq.\ (\ref{kappa1}) and perform a
numerical estimation of the asymmetry.

\boldmath
\subsection{$\Phi$ matrix in the lowest order ($\Phi_{L}^{\alpha\beta}$)}
\unboldmath

\begin{figure}[htb]
\centering
\includegraphics[width=5in]{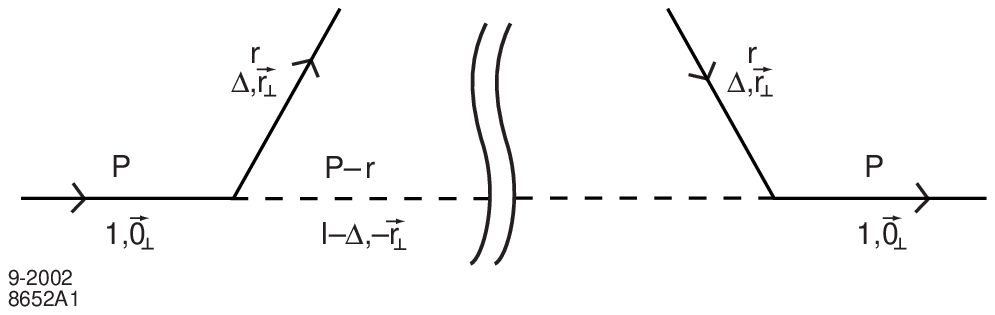}
\caption[*]{Diagram which gives the lowest order $\Phi$ (called
$\Phi_{L}^{\alpha\beta}$.} \label{PhiL}
\end{figure}

As indicated in Fig. \ref{PhiL} the initial proton has its
momentum given by $P^\mu = (P^+,P^-,{\bm
P}_{\perp})=(P^+,{M^2\over P^+},{\bm 0}_{\perp})$, and the final
diquark ${P'}^\mu = ({P'}^+,{P'}^-,{\bm
P}'_{\perp})=(P^+(1-\Delta), {{\lambda}^2+{\bm r}_{\perp}^2\over
P^+(1-\Delta)},{\bm r}_{\perp})$. We use the convention
$a^{\pm}=a^0\pm a^3$, $a\cdot b={1\over 2}(a^+b^-+a^-b^+)-{\bm
a}_{\perp}\cdot {\bm b}_{\perp}$.

We will first calculate the $\Phi$ matrix to lowest order
($\Phi_{L}^{\alpha\beta}$) in the quark-scalar diquark model
used in Ref.\ \cite{BHS}.  By calculation of Fig.\ \ref{PhiL} one
readily obtains
\begin{eqnarray}
\Phi_{L}^{\alpha\beta}&=&
ag^2{\Big[}{\bar{u}}(P,S){{\slsh{r}}+m\over r^2-m^2}{\Big]}^{\beta}
{\Big[}{{\slsh{r}}+m\over r^2-m^2}{u}(P,S){\Big]}^{\alpha}
\ {1\over P^+(1-\Delta)}
\label{d1}\\
&=&
ag^2{\Big[}{\bar{u}}(P,S)({\slsh{r}}+m){\Big]}^{\beta}
{\Big[}({\slsh{r}}+m){u}(P,S){\Big]}^{\alpha}
\ {1\over P^+(1-\Delta)} \
\nonumber\\
&&\times
\Bigl( {1\over
\Delta (M^2-{m^2+{\bm r}_{\perp}^{\, 2}\over \Delta}
-{\lambda^2+{\bm r}_{\perp}^{\, 2}\over 1-\Delta})}
{\Bigr)}^2
\ ,
\nonumber
\end{eqnarray}
with a constant $a=1/(2(2\pi)^3)$.  The normalization is fixed by the
condition
\begin{equation}
\int\ d\Delta\ d^2{\bm r}_{\perp}\ f_1(\Delta ,{\bm r}_{\perp})\ =\ 1\ .
\label{normb}
\end{equation}
In Eq.\ (\ref{d1}) we used the relation
\begin{eqnarray}
r^2-m^2&=&r^+(r^--{m^2+{\bm r}_{\perp}^{\, 2}\over r^+})
\ =\
r^+(P^--{\lambda^2+{\bm r}_{\perp}^{\, 2}\over (1-\Delta)P^+}
-{m^2+{\bm r}_{\perp}^{\, 2}\over r^+})
\nonumber\\
&=&
\Delta (M^2-{\lambda^2+{\bm r}_{\perp}^{\, 2}\over 1-\Delta}
-{m^2+{\bm r}_{\perp}^{\, 2}\over \Delta})
\ .
\label{d2}
\end{eqnarray}

This model is similar to the so-called spectator model (see e.g.\
Ref.\ \cite{Joao}), where in addition a vector diquark is included
and the coupling constant $g$ is treated as a form factor (in
order to guarantee convergence).  Of course, this can be assumed
in the present model calculation as well and will be discussed in
Section \ref{discussion}. Assuming real form factors, the
functions $f_{1T}^\perp$ and $h_1^\perp$ are strictly zero in the
spectator model.

\boldmath
\subsubsection{Calculation of $f_1(\Delta , {\bm r}_{\perp})$}
\unboldmath

For the calculation of the denominator of the asymmetry one needs to know the
function $f_1(\Delta , {\bm r}_{\perp})$,
which can be obtained from $\Phi^{\alpha\beta}$ given in Eq.\
(\ref{ad1}):
\begin{equation}
f_1(\Delta , {\bm r}_{\perp})=
{1\over 2}\sum_{\pm S}{1\over 2}\Phi^{\alpha\beta}(\gamma^+)^{\beta\alpha}\ .
\label{bd1}
\end{equation}
We now take $\Phi = \Phi_L$ and
for the numerator spinor contraction, we calculate
\begin{eqnarray}
&&{1\over 2}\sum_{\pm S}{\Big[}{\bar{u}}(P,S)({\slsh{r}}+m){\Big]}^{\beta}
{\Big[}({\slsh{r}}+m){u}(P,S){\Big]}^{\alpha}\ (\gamma^+)^{\beta\alpha}
\nonumber\\
&=&
{1\over 2}{\rm Tr}\Big[ ({\slsh{P}}+M)({\slsh{r}}+m)\gamma^+({\slsh{r}}+m)\Big]
\nonumber\\
&=&
2P^+\Big[ {\bm r}_{\perp}^{\, 2}+(\Delta M+m)^2\Big]\ .
\label{bd2}
\end{eqnarray}
Then, from Eqs.\ (\ref{d1}), (\ref{bd1}) and (\ref{bd2}), we arrive at
\begin{eqnarray}
f_1(\Delta , {\bm r}_{\perp})&=&
ag^2\ \Big[ {\bm r}_{\perp}^{\, 2}+(\Delta M+m)^2\Big]\
\ {1\over (1-\Delta)} \
\Bigl( {1\over
\Delta (M^2-{m^2+{\bm r}_{\perp}^{\, 2}\over \Delta}
-{\lambda^2+{\bm r}_{\perp}^{\, 2}\over 1-\Delta})}
{\Bigr)}^2
\nonumber\\
&=&{g^2\over 2(2\pi)^3}\
{(1-\Delta)\ \Big[ {\bm r}_{\perp}^{\, 2}+(\Delta M+m)^2\Big]\over
({\bm r}_{\perp}^{\, 2}+B)^2}
\ = \ C \frac{{\bm r}_{\perp}^{\, 2}+D}{({\bm r}_{\perp}^{\, 2}+B)^2},
\label{bd3}
\end{eqnarray}
where we define $C \equiv g^2 (1 - \Delta)/(2(2\pi)^3)$,
$D \equiv (\Delta M + m)^2$ and
\begin{equation}
B \ \equiv \ \Delta (1-\Delta)(-M^2+{m^2\over\Delta}+{\lambda^2\over
1-\Delta})\ .
\label{Bnote}
\end{equation}
Since we consider the proton state with mass $M$ as a bound state composed of
a quark with mass $m$ and a diquark with mass $\lambda$, the function
$B$ as given in Eq.\ (\ref{Bnote}) is always nonzero and positive.The integral
in Eq.\ (\ref{normb}) with $f_1(\Delta , {\bm r}_{\perp})$
given in Eq.\ (\ref{bd3}) can for instance be regulated by assuming a cutoff
in the invariant mass:
${\cal M}^2 = \sum_i {{\bm k}^2_{\perp i} +m^2_i \over x_i} < \Lambda^2$,
and the value of $g^2$ is adjusted to satisfy the normalization condition
Eq.\ (\ref{normb}) \cite{BHMS}.

\boldmath
\subsection{$\Phi$ matrix with final-state interaction
($\Phi_{F}^{\alpha\beta}$)} \unboldmath

\begin{figure}[htb]
\centering
\includegraphics[height=4in]{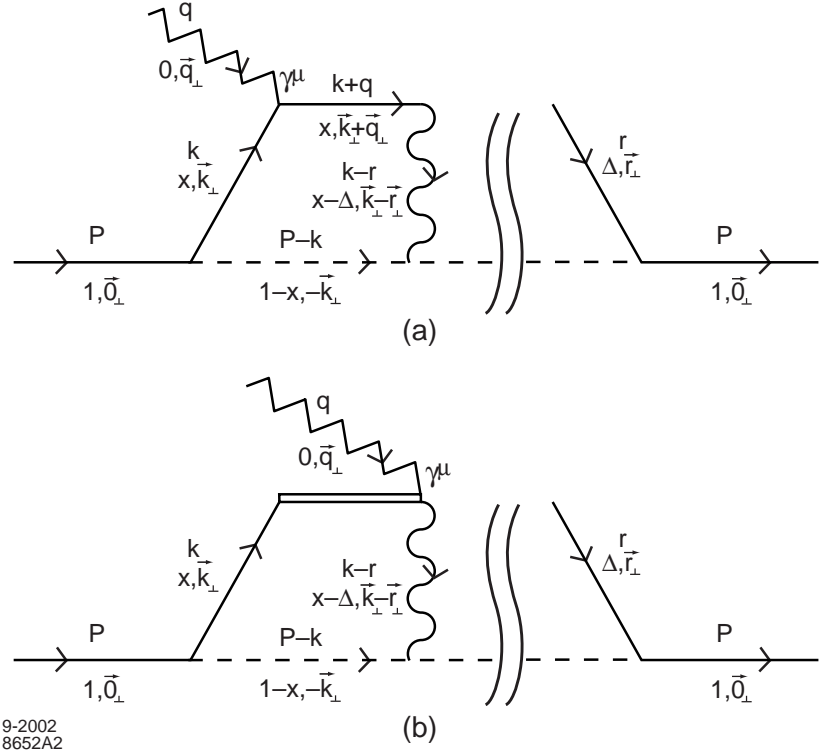}
\caption[*]{Diagrams which yield $\Phi$ with final-state
interaction ($\Phi_{F}^{\alpha\beta})$.} \label{PhiF}
\end{figure}

In order to obtain the $\Phi$ matrix with final-state interaction
(called $\Phi_{F}^{\alpha\beta}$), from which one can trivially
obtain the one with initial-state interaction, we calculate the
diagram given in Fig.\ \ref{PhiF}(b).  This is equal to the
diagram calculated by Ji and Yuan \cite{Ji-Yuan} to obtain nonzero
$f_{1T}^\perp$, starting from the formal gauge invariant
definition of this transverse momentum dependent distribution
function \cite{Collins-02,Belitsky}. In Fig.\ \ref{PhiF}(b) we
attached the virtual photon line to the later end of the eikonal
line in order to emphasize that the final-state interaction effect
has become an ingredient of the distribution functions of the
target proton.  In reality, the whole eikonal line should be
considered to be at the same point.

Defining $\Phi_{F}^{\alpha\beta}$ through Fig.\ \ref{PhiF}(b)
(in the Feynman gauge),
we have
\begin{eqnarray}
&&\Phi_{F}^{\alpha\beta}
\label{d3}\\
&=&iag^2\ e_1 e_2\ \int {d^4k\over (2\pi)^4}
\ {1\over 2}
\ {-P^+((1-x)+(1-\Delta))\over P^+(1-\Delta)}
\nonumber\\
&&\times
{2\over P^+((x-\Delta )+i\epsilon )}
\nonumber\\
&&\times
{{\Big[}{\bar{u}}(P,S)({\slsh{r}}+m){\Big]}^{\beta}
{\Big[}({\slsh{k}}+m){u}(P,S){\Big]}^{\alpha}\over
(k^2-m^2+i\epsilon )
((k-r)^2-\lambda_g^2+i\epsilon )((k-P)^2-\lambda^2+i\epsilon
)}\
{1\over r^2-m^2}
\nonumber\\
&&+\ ({\rm h.c.})
\nonumber\\
&=&-iag^2\ e_1 e_2\ \int {d^2{\bm{k}}_{\perp}\over 2(2\pi)^4}
\int P^+dx\ {1\over P^{+3}\ x\ (x-\Delta )\ (1-x)}\
{(-(1-x)-(1-\Delta))\over 2 P^+(1-\Delta)}
\nonumber\\
&&\times
{2\over ((x-\Delta )+i\epsilon )}\
{1\over
\Delta (M^2-{m^2+{\bm r}_{\perp}^{\, 2}\over \Delta}
-{\lambda^2+{\bm r}_{\perp}^{\, 2}\over 1-\Delta})}
\nonumber\\
&&\times
\int dk^-
{\Big[}{\bar{u}}(P,S)({\slsh{r}}+m){\Big]}^{\beta}
{\Big[}({\slsh{k}}+m){u}(P,S){\Big]}^{\alpha}
\nonumber\\
&&\times
{1\over
\left(k^--{(m^2+{\bm{k}}_{\perp}^2)-i\epsilon\over xP^+}\right)
\left((k^--r^-)-{(\lambda_g^2+({\bm{k}}_{\perp}-{\bm{r}}_{\perp})^2)
-i\epsilon\over (x-\Delta )P^+}\right)
\left((k^--P^-)+{(\lambda^2+{\bm{k}}_{\perp}^2)-i\epsilon\over
(1-x)P^+}\right)}
\nonumber\\
&&+\ ({\rm h.c.})\ ,
\nonumber
\end{eqnarray}
where we used $k^+=xP^+.$
The derivation of the starting formula of
Eq.\ (\ref{d3}) is given in the Appendix. This underlies the step from Fig.\
\ref{PhiF}(a) to Fig.\ \ref{PhiF}(b) and hence the step from Fig.\
\ref{LODY2} to Fig.\ \ref{LODY}.

For $\Phi_{F}^{\alpha\beta}$ in Eq.\ (\ref{d3}), we consider only
the contribution from the imaginary part of ${1\over ((x-\Delta
)+i\epsilon )}$, that is, the contribution from $-i\pi \delta
(x-\Delta )$. There is no contribution from the real part of
${1\over ((x-\Delta )+i\epsilon )}$, since the hermitian conjugate
term cancels it. Then, we have
\begin{eqnarray}
&&\Phi_{F}^{\alpha\beta}
\label{d3a}\\
&=&4\ (-iag^2\ e_1 e_2)\ \int {d^2{\bm{k}}_{\perp}\over 2(2\pi)^4}
\int P^+dx\ {1\over P^{+3}\ x\ (x-\Delta )\ (1-x)}\
{(-(1-x)-(1-\Delta))\over 2 P^+(1-\Delta)}
\nonumber\\
&&\times
\Big[ -i\pi \delta (x-\Delta )\Big]\
{1\over
\Delta (M^2-{m^2+{\bm r}_{\perp}^{\, 2}\over \Delta}
-{\lambda^2+{\bm r}_{\perp}^{\, 2}\over 1-\Delta})}
\nonumber\\
&&\times
\int dk^-
{\Big[}{\bar{u}}(P,S)({\slsh{r}}+m){\Big]}^{\beta}
{\Big[}({\slsh{k}}+m){u}(P,S){\Big]}^{\alpha}
\nonumber\\
&&\times
{1\over
\left(k^--{(m^2+{\bm{k}}_{\perp}^2)-i\epsilon\over xP^+}\right)
\left((k^--r^-)-{(\lambda_g^2+({\bm{k}}_{\perp}-{\bm{r}}_{\perp})^2)
-i\epsilon\over (x-\Delta )P^+}\right)
\left((k^--P^-)+{(\lambda^2+{\bm{k}}_{\perp}^2)-i\epsilon\over
(1-x)P^+}\right)}
\ .
\nonumber
\end{eqnarray}

When we perform the $k^-$ integration, we have
\begin{eqnarray}
&&\Phi_{F}^{\alpha\beta}
\label{d3b}\\
&=&4\ \pi\ ag^2\ e_1 e_2\ \int {d^2{\bm{k}}_{\perp}\over 2(2\pi)^4}
\int dx\ {1\over P^{+3}\ x\ (x-\Delta )\ (1-x)}\
{((1-x)+(1-\Delta))\over 2 (1-\Delta)}
\nonumber\\
&&\times
\delta (x-\Delta )\
{1\over
\Delta (M^2-{m^2+{\bm r}_{\perp}^{\, 2}\over \Delta}
-{\lambda^2+{\bm r}_{\perp}^{\, 2}\over 1-\Delta})}
\nonumber\\
&&\times
{\Big[}{\bar{u}}(P,S)({\slsh{r}}+m){\Big]}^{\beta}
{\Big[}({\slsh{k}}+m){u}(P,S){\Big]}^{\alpha}
\nonumber\\
&&\times
2\pi i\
{1\over
\left(P^--{(\lambda^2+{\bm{k}}_{\perp}^2)-i\epsilon\over
(1-x)P^+}
-{(m^2+{\bm{k}}_{\perp}^2)-i\epsilon\over xP^+}\right)
\left((
P^--{(\lambda^2+{\bm{k}}_{\perp}^2)-i\epsilon\over
(1-x)P^+}
-r^-)-{(\lambda_g^2+({\bm{k}}_{\perp}-{\bm{r}}_{\perp})^2)
-i\epsilon\over (x-\Delta )P^+}\right)}
\ .
\nonumber
\end{eqnarray}

When we perform the $x$ integration, we have
\begin{eqnarray}
&&\Phi_{F}^{\alpha\beta}
\label{d3c}\\
&=&-\ i\ ag^2\ e_1 e_2\ \int {d^2{\bm{k}}_{\perp}\over (2\pi)^2}
\ {1\over P^{+}\ \Delta\ (1-\Delta)}\
\nonumber\\
&&\times
{1\over
\Delta (M^2-{m^2+{\bm r}_{\perp}^{\, 2}\over \Delta}
-{\lambda^2+{\bm r}_{\perp}^{\, 2}\over 1-\Delta})}
\nonumber\\
&&\times
{\Big[}{\bar{u}}(P,S)({\slsh{r}}+m){\Big]}^{\beta}
{\Big[}({\slsh{k}}+m){u}(P,S){\Big]}^{\alpha}
\nonumber\\
&&\times
{1\over
\left( M^2-{(\lambda^2+{\bm{k}}_{\perp}^2)\over
(1-\Delta)}
-{(m^2+{\bm{k}}_{\perp}^2)-i\epsilon\over \Delta}\right)
\left( \lambda_g^2+({\bm{k}}_{\perp}-{\bm{r}}_{\perp})^2)
\right)}
\nonumber\\
&=&-\ i\ ag^2\ e_1 e_2\
{(1-\Delta) \over P^+({\bm r}_{\perp}^{\, 2}+B)}\
\int {d^2{\bm{k}}_{\perp}\over (2\pi)^2}
{{\Big[}{\bar{u}}(P,S)({\slsh{r}}+m){\Big]}^{\beta}
{\Big[}({\slsh{k}}+m){u}(P,S){\Big]}^{\alpha}
\over ({\bm k}_{\perp}^{\, 2}+B)\
(({\bm{k}}_{\perp}-{\bm{r}}_{\perp})^2+\lambda_g^2)}
\ ,
\nonumber
\end{eqnarray}
where $B$ is given in Eq.\ (\ref{Bnote}).

\boldmath
\subsubsection{Calculation of $f_{1T}^{\perp}(\Delta , {\bm r}_{\perp})$}
\unboldmath

One obtains $f_{1T}^{\perp}(\Delta , {\bm r}_{\perp})$
from $\Phi^{\alpha\beta}$ given in Eq.\ (\ref{ad1})
by extracting the proton
spin dependent part of $\Phi^{\alpha\beta}(\gamma^+)^{\beta\alpha}$:
\begin{equation}
\Phi^{\alpha\beta}(\gamma^+)^{\beta\alpha} =
2{\epsilon}^{ij}S_T^ir_{\perp}^j{f_{1T}^{\perp}\over M}\ ,
\label{cd1}
\end{equation}
where $\epsilon^{12}=+1$.

We now apply this to $\Phi_F$ and
for the numerator spinor contraction we calculate
\begin{eqnarray}
&&{\Big[}{\bar{u}}(P,S)({\slsh{r}}+m){\Big]}^{\beta}
{\Big[}({\slsh{k}}+m){u}(P,S){\Big]}^{\alpha}\ (\gamma^+)^{\beta\alpha}
\nonumber\\
&=&
{\rm Tr}\Big[ ({\slsh{P}}+M)({1\over 2}\gamma_5{\slsh{S}})
({\slsh{r}}+m)\gamma^+({\slsh{k}}+m)\Big]
\nonumber\\
&=&
(-{1\over 2})(-4i\epsilon^+_{\ \ \nu\rho\sigma})\Big[
mP^{\nu}S^{\rho}(k-r)^{\sigma}
+Mr^{\nu}S^{\rho}(k-r)^{\sigma}\Big]
\nonumber\\
&=&
-2iP^+(\Delta M+m)\epsilon^{ij}S_T^i(k_\perp-r_\perp)^j\ \ \
{\rm when}\ \ x=\Delta
\ ,
\label{cd2}
\end{eqnarray}
where we used $\epsilon^{0123}=+1$ and ${\rm
Tr}\Big[\gamma_5{\slsh{a}}{\slsh{b}}{\slsh{c}}{\slsh{d}}\Big]
=-4i\epsilon_{\mu\nu\rho\sigma}a^{\mu}b^{\nu}c^{\rho}d^{\sigma}$.

When we insert Eq.\ (\ref{cd2}) into Eq.\ (\ref{d3c}), we obtain
\begin{eqnarray}
&&\Phi_{F}^{\alpha\beta}(\gamma^+)^{\beta\alpha}\
[S\ {\rm dependent\ part}]
\label{fd3c}\\
&=&-\ i\ ag^2\ e_1 e_2\
{(1-\Delta) \over P^+({\bm r}_{\perp}^{\, 2}+B)}\
\int {d^2{\bm{k}}_{\perp}\over (2\pi)^2}
{-2iP^+(\Delta M+m)\epsilon^{ij}S_T^i(k_\perp-r_\perp)^j
\over ({\bm k}_{\perp}^{\, 2}+B)\
(({\bm{k}}_{\perp}-{\bm{r}}_{\perp})^2+\lambda_g^2)}
\nonumber\\
&=&-\ 2\ ag^2\ e_1 e_2
\ {(\Delta M + m)(1-\Delta) \over ({\bm r}_{\perp}^{\, 2}+B)}\
\epsilon^{ij}S_T^i\
\int {d^2{\bm{k}}_{\perp}\over (2\pi)^2}
{(k_\perp-r_\perp)^j
\over ({\bm k}_{\perp}^{\, 2}+B)\
(({\bm{k}}_{\perp}-{\bm{r}}_{\perp})^2+\lambda_g^2)}
\ .
\nonumber
\end{eqnarray}

Then, from Eqs.\ (\ref{cd1}) and (\ref{fd3c}) we get
\begin{equation}
f_{1T}^{\perp}(\Delta , {\bm r}_{\perp})\ =\
-\ {1\over 4\pi}\
ag^2\ e_1 e_2\
\ {M(\Delta M + m)(1-\Delta) \over ({\bm r}_{\perp}^{\, 2}+B)}\
{1\over {\bm r}_{\perp}^{\, 2}}
\ln
\left(\frac{{\bm r}_{\perp}^{\, 2} + B}{B}\right)
\ .
\label{cd3}
\end{equation}

{}From Eq.\ (\ref{ad1}) we find that in terms of $f_{1}$ and $f_{1T}^{\perp}$
the single-spin asymmetry transverse to the production plane in the SIDIS
is given by (with ${\bm S}_T = S_T^2\, {\hat y}$)
\begin{equation}
{\cal P}_y\ S_T^2 \ =\
{{\epsilon}^{+-}_{\ \ \ \,\, ij}\,\, r_{\perp}^i\, S_T^j\,
f_{1T}^{\perp}(\Delta , {\bm r}_{\perp})\over
2\, M\, f_{1}(\Delta , {\bm r}_{\perp})}
\ =\
-\ {r_{\perp}^1\over M}\ {f_{1T}^{\perp}(\Delta , {\bm r}_{\perp})\over
f_{1}(\Delta , {\bm r}_{\perp})}\ S_T^2 \ .
\label{py1}
\end{equation}
Then, using the results in Eqs.\ (\ref{bd3}) and (\ref{cd3}), we get
\begin{equation}
{\cal P}_y\ =\
{e_1 e_2\over 4\pi}\
{(\Delta M + m)\ r_{\perp}^1 \over {\bm r}_{\perp}^{\, 2} + (\Delta M + m)^2}\
{{\bm r}_{\perp}^{\, 2} + B\over {\bm r}_{\perp}^{\, 2}}\
\ln
\left(\frac{{\bm r}_{\perp}^{\, 2} + B}{B}\right)
\ ,
\label{py2}
\end{equation}
which agrees with Eq.\ (21) of Ref.\ \cite{BHS}.

\boldmath
\subsubsection{Calculation of $h_{1}^{\perp}(\Delta , {\bm r}_{\perp})$}
\unboldmath

Similarly, one obtains $h_{1}^{\perp}(\Delta , {\bm r}_{\perp})$
from $\Phi^{\alpha\beta}$ given in Eq.\ (\ref{ad1}) by extracting
the proton spin independent part of
$\Phi^{\alpha\beta}(\sigma^{i+})^{\beta\alpha}$:
\begin{equation}
\Phi^{\alpha\beta}(\sigma^{i+})^{\beta\alpha} =
2r_{\perp}^i{h_{1}^{\perp}\over M}\ ,
\label{ed1}
\end{equation}
where $\sigma^{\mu\nu}={i\over 2}[\gamma^\mu , \gamma^\nu ]$.

We again apply this to $\Phi_F$ and for the numerator spinor
contraction, we obtain
\begin{eqnarray}
&&{1\over 2}\sum_{\pm S}{\Big[}{\bar{u}}(P,S)({\slsh{r}}+m){\Big]}^{\beta}
{\Big[}({\slsh{k}}+m){u}(P,S){\Big]}^{\alpha}\ (\sigma^{i+})^{\beta\alpha}
\nonumber\\
&=&
{1\over 2}{\rm Tr}\Big[ ({\slsh{P}}+M)
({\slsh{r}}+m)({i\over 2}(\gamma^i\gamma^+
-\gamma^+\gamma^i))({\slsh{k}}+m)\Big]
\nonumber\\
&=&
-2i\ P^+\Big[ M(\Delta k^i-xr^i)+m(k-r)^i\Big]
\nonumber\\
&=&
-2i\ P^+(\Delta M+m)\ (k_\perp-r_\perp)^i\ \ \
{\rm when}\ \ x=\Delta
\ .
\label{fcd2}
\end{eqnarray}

Then, from Eqs.\ (\ref{d3c}), (\ref{ed1}) and (\ref{fcd2}) we obtain
\begin{equation}
h_{1}^{\perp}(\Delta , {\bm r}_{\perp})\ =\
-\ {1\over 4\pi}\
ag^2\ e_1 e_2\
\ {M(\Delta M + m)(1-\Delta) \over ({\bm r}_{\perp}^{\, 2}+B)}\
{1\over {\bm r}_{\perp}^{\, 2}}
\ln
\left(\frac{{\bm r}_{\perp}^{\, 2} + B}{B}\right)
\ .
\label{cd3a}
\end{equation}

Thus, from Eqs.\ (\ref{cd3}) and (\ref{cd3a}) we find the relation
\begin{equation}
f_{1T}^{\perp}(\Delta , {\bm r}_{\perp})\ =\ h_{1}^{\perp}(\Delta
, {\bm r}_{\perp})\ .
\label{ae1}
\end{equation}
We note that the equality Eq.\ (\ref{ae1}) is a special property of the
quark-scalar diquark model.

We can write $f_{1T}^{\perp}$ and $h_{1}^{\perp}$ given in
Eqs.\ (\ref{cd3}) and (\ref{cd3a}) schematically as
\beq
f_{1T}^{\perp}(\Delta , {\bm r}_{\perp})=
h_{1}^{\perp}(\Delta , {\bm r}_{\perp}) = \frac{A}{ {\bm
r}_{\perp}^{\, 2}\ ({\bm r}_{\perp}^{\, 2}+B)} \ln
\left(\frac{{\bm r}_{\perp}^{\, 2} + B}{B}\right),
\label{h1perpschem} \eeq
with $B$ as given in Eq.\ (\ref{Bnote}) and
\begin{equation}
A\ =\ {g^2\over 2(2\pi)^3}\ \Bigl( -\ {e_1 e_2\over 4\pi}\Bigr)\
{M\ (\Delta M + m)} (1-\Delta)\ .
\label{Anote}
\end{equation}
We have the same formulas for $\overline f_{1T}^{\perp}$ and
$\overline h_1^{\perp}$ with $\Delta, {\bm r}_{\perp},
A, B$ replaced by $\bar \Delta, {\bm {\bar r}}_{\perp},
\bar A, \bar B$.

We note that we obtained $f_{1T}^{\perp}$ and $h_{1}^{\perp}$ in
Eq.\ (\ref{h1perpschem}) from the final-state interaction diagram
shown in Fig. \ref{PhiF}(b). These are the functions relevant for
semi-inclusive DIS \cite{BHS}. The functions arising from
initial-state interactions have an overall minus sign compared to
those in Eq.\ (\ref{h1perpschem}), as pointed out by
\cite{Collins-02} and confirmed in \cite{BHS2}. However,
$\overline f_{1T}^{\perp}$ and $\overline h_1^{\perp}$ also have
this property, therefore, the asymmetry factor $\nu$ given in Eq.\
(\ref{kappa1}) is in fact independent of whether we use
$h_{1}^{\perp}$ and $\overline h_1^{\perp}$ from initial- or
final-state interactions.

\boldmath
\subsection{The $\cos 2\phi$ asymmetry}
\unboldmath

We now consider the convolution terms in the numerator and denominator of the
analyzing power $\nu$ of the asymmetry (Eq.\ (\ref{kappa1})):
\ba
F & \equiv & {\cal F}\left[\left(2\,\bm{\hat h}\!\cdot \!
\bm p_\perp^{}\,\,\bm{\hat h}\!\cdot \! \bm k_\perp^{}\,
                    -\,\bm p_\perp^{}\!\cdot \! \bm k_\perp^{}\,\right)
                   h_1^{\perp}\overline h_1^{\perp} \right] \nn \\[2 mm]
& = & \int d^2\bm p_\perp^{} d^2\bm k_\perp^{}
\delta^2 (\bm p_\perp^{}+\bm k_\perp^{}-\bm
q_\perp^{})  \left(2\,\bm{\hat h}\!\cdot \!
\bm p_\perp^{}\,\bm{\hat h}\!\cdot \! \bm k_\perp^{}
                    -\bm p_\perp^{}\!\cdot \! \bm k_\perp^{}\right)
                   h_1^{\perp}(\Delta,\bm{p}_\perp^2)
\overline h_1^{\perp}(\bar \Delta,\bm{k}_\perp^2), \nn\\[2 mm]
G & \equiv & {\cal F}\left[f_1 \overline f_1 \right] \nn \\[2 mm]
& = & \int d^2\bm p_\perp^{}\; d^2\bm k_\perp^{}\;
\delta^2 (\bm p_\perp^{}+\bm k_\perp^{}-\bm
q_\perp^{}) f_1(\Delta,\bm{p}_\perp^2)
\overline f_1(\bar \Delta,\bm{k}_\perp^2),
\label{conv1}
\ea
where we left out the flavor indices. With these definitions we can write
\beq
\nu = \frac{2}{M_1 M_2} \frac{\sum_{a,\bar a} e_a^2 \, F_a}{\sum_{a,\bar a}
e_a^2 \, G_a}.
\eeq
We will insert the schematic form Eq.\ (\ref{bd3})
for $f_1$ and $\overline f_1$
and Eq.\ (\ref{h1perpschem}) for $h_1^{\perp}$ and $\overline h_1^{\perp}$.

We first rewrite the denominator term $G$:
\beq
G =  \int \frac{d^2 \bm b_\perp^{}}{(2\pi)^2} \, \exp \left(-i \bm b_\perp^{}
\cdot \bm q_\perp^{}\right) \, \tilde f_1 (\Delta,\bm b_\perp^{2}) \,
\tilde{\overline f}{}_1(\bar \Delta,\bm b_\perp^{2}),
\eeq
where we have defined the Fourier transform of $f_1(\Delta,\bm k_\perp^{2})$
\ba
\tilde f_1 (\Delta,\bm b_\perp^{2}) & \equiv & \int \, d^2\bm p_\perp^{}\,
\exp \left(i \bm b_\perp^{} \cdot \bm p_\perp^{}\right)
\, f_1(\Delta,\bm p_\perp^{2}) \nn\\
& = & 2\pi C \left(K_0(\sqrt{B}b) + {(D-B)\over 2{\sqrt{B}}} b
K_1(\sqrt{B}b)\right),
\ea
where $b \equiv |\bm b_\perp^{}|$, and similarly for $\overline f_1$.
Thus, we obtain the exact expression for $G$:
\ba
G & = & 2\pi C\bar C \int_0^\infty d b \, b \,
J_0(b|\bm{q}_\perp|) \, \left(K_0(\sqrt{B}b) + {(D-B)\over 2{\sqrt{B}}} b
K_1(\sqrt{B}b)\right)\nn
\\
& & \times
\left(K_0(\sqrt{\bar B}b) + {(\bar D-\bar B)\over 2{\sqrt{\bar B}}} b
K_1(\sqrt{\bar B}b)\right).
\ea

Obtaining such an exact expression for $F$ is much more difficult (if possible
at all), hence we will express $F$ in a form amenable to
numerical evaluation. We first write
\beq
F = - \int_0^\infty \frac{d b}{2\pi} \, b \,
J_2(b|\bm{q}_\perp|) \, \tilde{h}_1^\perp(\Delta,b) \,
\tilde{\overline h}{}_1^\perp(\bar \Delta, b),
\eeq
where we have defined the Fourier transform
\ba
\tilde{h}_1^\perp(\Delta,b) & \equiv & \int \, d^2\bm p_\perp^{}\,
\exp\left(i \bm b_\perp^{} \cdot \bm p_\perp^{}\right)\, \frac{\bm
b_\perp^{} \cdot \bm p_\perp^{}}{b} \,
h_1^\perp(\Delta,\bm p_\perp^{}) \nn\\
& = & 2\pi i A \int d p J_1(b p) \ln \left(\frac{p^2+B}{B}\right)
\frac{1}{p^2+B}, \ea where $p \equiv |\bm p_\perp^{}|$. This can
be approximated from below by expanding \beq \ln
\left(\frac{p^2+B}{B}\right) = \left(\frac{p^2}{p^2+B}\right)+ {1
\over 2} \left(\frac{p^2}{p^2+B}\right)^2+ {1\over
3}\left(\frac{p^2}{p^2+B}\right)^3 + \cdots \ . \eeq For each term
an exact Fourier transform expression can be obtained in terms of
$K_i$ functions. Keeping only the first term will lead for
instance to
\beq
\tilde{h}_1^\perp(\Delta,b) \simorder - \frac{i
\pi A}{\sqrt{B}} \left( K_1(\sqrt{B}b) - {1 \over 2} b \sqrt{B}
\left[K_0(\sqrt{B}b)+K_2(\sqrt{B}b) \right]\right),
\label{h1perpapprox}
\eeq
which is roughly a factor of 2 too small
compared to the numerical evaluation without approximation. Eq.\
(\ref{h1perpapprox}) leads to an asymmetry with approximately the
right shape, but about a factor of 4 smaller in magnitude. This
discrepancy can be reduced by taking further terms in the Taylor
expansion into account.

We will now investigate the obtained expressions for $F$ and $G$
by a numerical evaluation. In order to simplify the numerical
calculation somewhat (since no absolute prediction can be made at
this stage, because the overall magnitude of $A$ and $\bar A$ are
not known), we assume the situation of equal hadron masses
($M_1=M_2=M$) and take momentum fractions such that $B=\bar B$ and
$D=\bar D$. This results in the following expressions, after
expressing all dimensionful two-vectors in units of $\sqrt{B}$,
i.e.\ rescaling $\bm b_\perp^{} \to \sqrt{B}\bm b_\perp^{}$ and
$\bm q_\perp^{} \to \bm q_\perp^{}/\sqrt{B}$ (idem for $\bm
p_\perp^{}$ and $\bm k_\perp^{}$), \ba F & = & \frac{2\pi A \bar
A}{B^2} \int \, db\, b\, J_2(b|\bm{q}_\perp|) \, \left( \int d p
\, J_1(b p) \, \ln \left(p^2+1 \right)\frac{1}{p^2+1} \right)^2,
\nn \\[2 mm]
G & = & \frac{2\pi C \bar C}{B} \int \, db\, b\, J_0(b|\bm{q}_\perp|) \,
\left( \int d p \, p \, J_0(b p) \frac{p^2+{D/B}}{(p^2+1)^2} \right)^2.
\ea

Next we make some generic choices for the various parameters.
We take $M=0.94 \, \text{GeV}, m=0.3 \, \text{GeV}, \lambda = 0.8 \,
\text{GeV}$ and $D/B=4$, which implies that $\Delta \approx 0.2 $ or
$0.5$ and idem for $\bar \Delta$.

\begin{figure}[htb]
\centering
\includegraphics[height=2in]{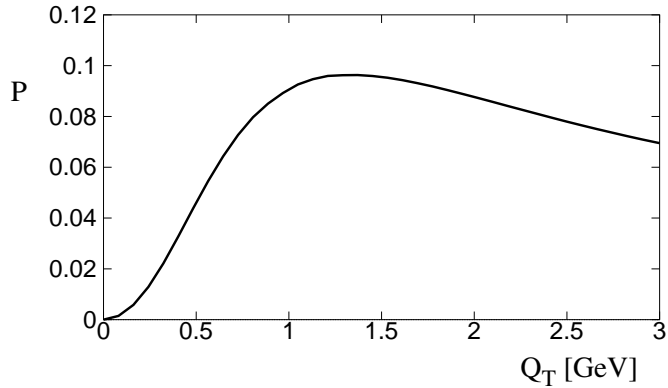}
\caption[*]{Numerical result for $P \equiv B C \bar C F /(A \bar A
G)$, using $M=0.94 \, \text{GeV}, m=0.3 \, \text{GeV}, \lambda =
0.8 \, \text{GeV}$ and $\Delta=\bar \Delta = 0.2$.}
\label{DYcos2phi}
\end{figure}

Figure\ \ref{DYcos2phi} displays the quantity $P \equiv B C \bar C
F /(A \bar A G)$ as function of $|\bm q_\perp^{}|$ in GeV (using
$\Delta=0.2$, $\sqrt{B} \approx 0.24$ GeV). The quantity $P$ still
has to be related to $\nu$ which cannot be done without further
assumptions. First of all, we will assume $u$ quark dominance,
which yields $\nu \approx 2 F_u/(M_1 M_2 G_u)$. Next we will use
some results obtained in Ref.\ \cite{Boer}, where the same
asymmetry $\nu$ was investigated and the following form was
assumed (based on very general arguments and the simple model
result of Ref.\ \cite{Collins-93b}):
\beq
\frac{h_1^\perp{}(\Delta,
\bm{p}_\perp^2)}{f_1(\Delta,\bm{p}_\perp^2)} = c_H \frac{M_C
M_H}{\bm{p}_\perp^2+ M_C^2}, \label{hoverf1}
\eeq
where $M_H$ is
the mass of the hadron and $c_H$ and $M_C$ were used as fitting
parameters. The values $c_H \approx 1$ and $M_C \approx 2$ were
obtained from fitting the 194 GeV data of the NA10 Collaboration,
by considering the case of one dominant quark flavor contribution.
In the present model calculation the ratio takes the form
\beq
\frac{h_1^\perp{}(\Delta,
\bm{p}_\perp^2)}{f_1(\Delta,\bm{p}_\perp^2)} = \frac{A}{C}
\frac{1}{\bm{p}_\perp^2} \frac{\bm{p}_\perp^2+B}{\bm{p}_\perp^2+D}
\ln \left( \frac{\bm{p}_\perp^2+B}{B}\right).
\label{hoverf1a}
\eeq
Unfortunately this shape is very different from Eq.\
(\ref{hoverf1}) for the choices of $B$ and $D$ made earlier
($\Delta \approx 0.2 $ such that $B \approx 1/16$ and $D \approx
1/4$). Although both forms have similar large $\bm{p}_\perp^2$
behavior, it is mostly the small $\bm{p}_\perp^2$ behavior that is
relevant. By comparing the curves resulting from Eq.\
(\ref{hoverf1}) (with $c_H=1$, $M_C = 2 M_H = 2$ GeV) and Eq.\
(\ref{hoverf1a}) (with $D = 4 B = 1/4$), one may expect $0.1
\simordertwo A/C \simordertwo 0.5$, which then implies that $2
\simordertwo |e_1 e_2| \simordertwo 12$ (incidentally this matches
the value of $|e_1 e_2| \approx 5$, which was used for the
numerical estimation in Ref.\ \cite{BHS}). This range of values
may then be used for crude estimates of asymmetries containing the
function $h_1^\perp$ for a quark inside a proton, or equivalently
an anti-quark inside an anti-proton. For a quark inside an
anti-proton, or equivalently an anti-quark inside a proton, the
overall prefactor is expected to be smaller. So if we restrict
ourselves to $p \bar p $ collisions here and take $A/C=\bar A/\bar
C = 0.3$, then one obtains as a very crude estimate: $\nu \approx
2 F/(M_1 M_2 G) = 2 A^2 P/(B C^2) \approx 3 P$, which means that
the maximum of $\nu$ is on the order of 30\%. As said this is a
very crude estimate and many assumptions went in. It cannot be
viewed as more than an order of magnitude estimate, but we think
it is an encouraging result.

\subsection{\label{discussion}Discussion}

In this section we give a more general discussion of the
qualitative features of the asymmetry, in particular its
$Q_\perp^2$ dependence. It may be good to note that the starting
point of the calculation, that is, the factorized description of
the asymmetry, requires that $Q_\perp^2 \ll Q^2$, such that for
large values of $Q_\perp^2$ the asymmetry is not appropriately
described by the above formulas. At $Q_\perp^2 \sim Q^2$, the
perturbative corrections will be the dominant source of an
asymmetry.

In order to obtain the general $Q_\perp^2$ dependence of the
asymmetry for small and large $Q_\perp^2$, we start with the
original convolution expression for $F$ (the first line of Eq.\
(\ref{conv1})). After multiplication by a trivial factor
$Q_\perp^2/Q_\perp^2$ and using the $\bm{k}_\perp$ integration to
eliminate the delta function, we shift the integration variable
$\bm{p}_\perp \to \bm{p}_\perp'=\bm{p}_\perp^{} -{\scriptstyle
{1\over 2}}\bm{q}_\perp^{}$, to arrive at
\ba
F & = & \frac{A \bar
A}{Q_\perp^2} \int d ^2\bm p_\perp' \left(\frac{Q_\perp^4}{4} +
{\bm p_\perp'}^2 Q_\perp^2 - 2 (\bm q_\perp^{} \cdot \bm
p_\perp')^2 \right) \frac{1}{ (\bm p_\perp' + {\scriptstyle
{1\over 2}}\bm{q}_\perp^{})^2 (\bm p_\perp' - {\scriptstyle
{1\over 2}}\bm{q}_\perp^{})^2 }
\nn \\[2 mm] & & \times
\frac{1}{ \left((\bm p_\perp' + {\scriptstyle {1\over 2}}\bm{q}_\perp^{})^2
+ B \right)
\left( (\bm p_\perp' -
{\scriptstyle {1\over 2}}\bm{q}_\perp^{})^2 + \bar B \right) }
\nn \\[2 mm]
& & \times \ln \left(\frac{(\bm p_\perp' + {\scriptstyle {1\over
2}}\bm{q}_\perp^{})^2 + B}{B}\right) \ln \left(\frac{(\bm p_\perp' -
{\scriptstyle {1\over 2}}\bm{q}_\perp^{})^2 + \bar B}{\bar B}\right).
\label{Fprime}
\ea
In case $B = \bar B$ (which, as said, means equal masses of the initial
hadrons and equal light-cone momentum fractions of the quark and
antiquark), then one can perform symmetric integration to reduce
\beq \left(\frac{Q_\perp^4}{4} + {\bm p_\perp'}^{2} Q_\perp^2 - 2
(\bm q_\perp^{} \cdot \bm p_\perp')^2 \right) \to \frac{Q_\perp^4}{4}. \eeq

For small $Q_\perp$ one can ignore the
$\pm {\scriptstyle {1\over 2}}\bm{q}_\perp^{}$ terms
in the denominators and $\ln$ terms of the expression
Eq.\ (\ref{Fprime}), such that symmetric integration is appropriate and one
can immediately conclude that $F \sim Q_\perp^2$.

Next we turn to the denominator of
$\nu$ (Eq.\ (\ref{kappa1})) which is given by
\beq
G = C \bar C \int d ^2\bm p_\perp' \frac{(\bm p_\perp' +
{\scriptstyle {1\over 2}}\bm{q}_\perp^{})^2 + D}{\left((\bm p_\perp' +
{\scriptstyle {1\over
2}}\bm{q}_\perp^{})^2 + B\right)^2} \frac{(\bm p_\perp' - {\scriptstyle {1\over
2}}\bm{q}_\perp^{})^2 + \bar D}{\left((\bm p_\perp' - {\scriptstyle {1\over
2}}\bm{q}_\perp^{})^2 + \bar B\right)^2}.
\eeq
At $Q_\perp^2=0 $, $G \neq 0$, hence the asymmetry ($\nu \sim F/G$) vanishes.

For large $Q_\perp$ the obtained model expressions do not yield an
accurate description, although one does obtain a power law
fall-off (see below), as one also would expect from perturbative
QCD (which determines the large transverse momentum region). The
point is that one runs into convergence problems. This applies for
instance to the integral over $f_1$ (Eq.\ (\ref{normb})). Also
$h_1^\perp$ does not fall off fast enough at large $Q_\perp^2$ to
guarantee convergence of certain $Q_\perp^2$-weighted and
integrated asymmetries (such as investigated in Refs.\
\cite{Boer,Boer3}). Although one obtains a finite result for
\beq
\int d^2 {\bm r}_{\perp} h_{1}^{\perp}(\Delta , {\bm
r}_{\perp}^{\, 2}) = \frac{\pi^3 A }{6 B},
\eeq
this is however
not the object one encounters in the $\cos 2\phi$ asymmetry, nor
in $Q_\perp^2$-weighted and integrated asymmetries. Rather one
encounters in such weighted asymmetries the quantity \beq \int d^2
{\bm r}_{\perp} {\bm r}_{\perp}^{\, 2} h_{1}^{\perp}(\Delta , {\bm
r}_{\perp}^{\, 2}), \eeq which diverges in the quark-scalar
diquark model employed here.

Therefore, one often assumes that the proton-quark-diquark
coupling constants $g$ are in fact form factors, see for instance
Ref.\ \cite{Joao}. In the present quark-scalar diquark model
calculation no such form factors are included (although the use of
a regulator is implicitly assumed), because it would add another
complication to the evaluation of the asymmetry and more
importantly, in separating the perturbatively generated $\cos
2\phi$ asymmetry (which is only relevant at large $Q_\perp^2$),
from the nonperturbative contribution $\sim h_1^\perp \times
\overline h_1^\perp$, one has to impose an upper cut-off on the
$Q_\perp^2$ range anyway. Our interest here is not in the specific
fall-off of the asymmetry at large $Q_\perp^2$, but rather in the
moderate $Q_\perp^2$ region, where the contribution to the
asymmetry arising from initial-state interactions is maximal.

For large $Q_\perp^2$ one concludes from the above expressions
(after including a regulator to insure convergence, e.g.\ a
transverse momentum fall-off in $g$), that $F, G$ and $\nu \sim
F/G$ decrease for large $Q_\perp^2$. To see this in more detail,
we will approximate $F/G$ crudely by setting $\bm p_\perp^{'\, 2}
=0$ in the denominators and by ignoring $B, \bar B, D, \bar D$ and
the $\ln$ terms altogether. In this way we obtain for the large
$Q_\perp$ behavior of the ratio
\beq
\frac{F}{G} \sim
\frac{1}{Q_\perp^2},
\eeq
i.e.\ the asymmetry indeed falls off for
large $Q_\perp^2$. Since at small $Q_\perp^2$ the ratio $F/G$
grows as $Q_\perp^2$, there has to be a turn-over in $\nu$ as
function of $Q_\perp^2$, which has not yet been observed in
experiment, but is clearly seen in the model calculation reported
here.

We want to emphasize that the quantities which determine
the magnitude (and width) of the asymmetry $\nu$ are the same as those
appearing in the expression for the single-spin asymmetry proportional to $h_1
\times h_1^\perp$ and in the context of the model also for the single-spin
asymmetries discussed in Refs.\ \cite{BHS,BHS2} that depend on the Sivers
distribution function. Thus, the parametric dependencies
of these asymmetries can in principle be checked for consistency, in order
to see whether it is at least consistent to assume that the asymmetries
are generated by the same underlying mechanism. An example of such a
comparison was given in Ref.\ \cite{Boer5}.

One final comment is on the $Q^2$ scale. The model does not
produce a dependence on that scale and the $Q^2$ dependence of
transverse momentum dependent asymmetries is a notoriously
difficult problem (cf.\ e.g.\ \cite{DBnpb01,Henneman}). Due to the
lack of knowledge of this $Q^2$ dependence, we can only expect
that the asymmetry expression and the result from the model
calculation should apply to the same $Q^2$ range
($Q^2=m_{\gamma^*}^2= (4 - 12 \, \text{GeV})^2$) as that of the
existing Drell-Yan data, from which we used fitting results.

\section{Conclusions}

In this paper we have studied the $\cos 2 \phi$ distribution in
{\it unpolarized} Drell-Yan lepton pair production within the
context of a quark--scalar diquark model for the proton including
an initial-state gluon interaction. Such initial- or final-state
interactions lead to the appearance of (naive) $T$-odd
distribution functions, such as the Sivers effect function
$f^\perp_{1T}(x,\bm{p}_\perp^2)$ and its chiral-odd partner
$h_1^\perp(x,\bm{p}_\perp^2)$ \cite{Goldstein:2002vv,Ji-Yuan}. We
calculated those functions in the quark-scalar diquark model and
found that they are equal in this model. Even though this equality
is not expected to be satisfied generally in other models, this
result does show that $f^\perp_{1T}(x,\bm{p}_\perp^2)$ and
$h_1^\perp(x,\bm{p}_\perp^2)$ are closely related and are expected
to have similar magnitudes in general. With the model expressions
for $f_1$ and $h_1^\perp$ we were able to write down an expression
for the analyzing power $\nu$ of the $\cos 2 \phi$ asymmetry in
the unpolarized Drell-Yan process. Under the assumption of $u$
quark dominance and by using fitting results of Ref.\ \cite{Boer},
we have given a numerical estimation of the asymmetry for the $
p\, \bar p \to \ell^+ \ell^- X$ process. As an order of magnitude
estimate we obtained for the maximum of $\nu$ a value of $30\%$.
Despite the considerable uncertainty it is clear that based on
this model calculation the $\cos 2 \phi$ asymmetry can be of the
same order of magnitude in $ p\, \bar p \to \ell^+ \ell^- X$ as
experimentally measured results in $\pi^- \, N \rightarrow \mu^+
\, \mu^-\, X$ (in the same range of $Q^2$ values). It is natural
to expect that the asymmetry in $ p\, p \to \ell^+ \ell^- X$ will
be considerably smaller, but may still be expected to be on the
percent level.

Since the same mechanism (initial/final-state interactions) leads
to nonzero functions $f^\perp_{1T}(x,\bm{p}_\perp^2)$ and
$h_1^\perp(x,\bm{p}_\perp^2)$, it is clear that the single-spin
asymmetries in the SIDIS and the Drell-Yan process are closely
related to the $\cos 2 \phi$ asymmetry of the unpolarized
Drell-Yan process. Since the width and the magnitude of these
asymmetries are determined by the same parameters in the model,
one can relate the asymmetries and this may be tested by
experimental data. All this provides new insight into the role of
quark and gluon orbital angular momentum as well as of initial-
and final-state gluon exchange interactions in hard QCD processes.

\section*{Acknowledgements}

We wish to thank John Collins and Piet Mulders for helpful discussions.
The research of D.B.\ has been made possible by financial support from the
Royal Netherlands Academy of Arts and Sciences.

\section*{Appendix: Derivation of Eq.\ (\ref{d3})}

We present the derivation (based on Ref.\ \cite{Collins-02})
of the starting formula of Eq.\ (\ref{d3}):
\begin{eqnarray}
&&{\bar{u}}(q+r)\gamma^-{1\over {\slsh{k}}+{\slsh{q}}-M}\gamma^{\mu}
\nonumber\\
&=&
{1\over (k-r)^++i\epsilon}\
{\bar{u}}(q+r)\gamma^-(k-r)^+{1\over {\slsh{k}}+{\slsh{q}}-M}\gamma^{\mu}
\nonumber\\
&\simeq&
{1\over (k-r)^++i\epsilon}\
{\bar{u}}(q+r)\
2\, ({\slsh{k}}-{\slsh{r}})\
{1\over {\slsh{k}}+{\slsh{q}}-M}\gamma^{\mu}
\nonumber\\
&=&
{2\over (k-r)^++i\epsilon}\
{\bar{u}}(q+r)
\Big[ ({\slsh{k}}+{\slsh{q}}-M)-({\slsh{q}}+{\slsh{r}}-M)\Big]
{1\over {\slsh{k}}+{\slsh{q}}-M}\gamma^{\mu}
\nonumber\\
&=&
{2\over (k-r)^++i\epsilon}\
{\bar{u}}(q+r)
({\slsh{k}}+{\slsh{q}}-M){1\over {\slsh{k}}+{\slsh{q}}-M}\gamma^{\mu}
\nonumber\\
&=&
{2\over (k-r)^++i\epsilon}\
{\bar{u}}(q+r)\gamma^{\mu}
\ ,
\label{ap1}
\end{eqnarray}
where we used the equation of motion
${\bar{u}}(q+r)({\slsh{q}}+{\slsh{r}}-M)=0$ in the fourth line.
In the above, $2 / ((k-r)^++i\epsilon)$ is the eikonal propagator.

Before going from the second to the third line in Eq. (\ref{ap1}),
we deformed the contour of integration to the upper infinity in
the complex $(k-r)^+$ plane so that $|(k-r)^+|\ \gg \
|(k-r)^-|,|(k-r)^i|$ is satisfied along the new contour. We note
that what we deformed is the line along which the $(k-r)^+$
integration is performed, and the pole position of
$(k-r)^++i\epsilon = 0$ (at which we compute the value of the
residue) is not influenced by this deformation.

\end{document}